\documentclass[final]{aipproc}\layoutstyle{6x9}\begin{document}
\title{SPINLESS SALPETER EQUATION\\Some (Semi-) Analytical Approaches}
\author{Wolfgang Lucha}{address={Institute for High Energy Physics, Austrian
Academy of Sciences,\\Nikolsdorfergasse 18, A-1050 Vienna, Austria\\E-mail:
wolfgang.lucha@oeaw.ac.at}}\author{Franz F.~Sch\"oberl}{address={Institute
for Theoretical Physics, University of Vienna,\\Boltzmanngasse 5, A-1090
Vienna, Austria\\E-mail: franz.schoeberl@univie.ac.at}}\maketitle

The eigenvalue equation of a semirelativistic Hamiltonian composed of the
relativistic kinetic term of spin-0 particles and static interactions is
called spinless Salpeter equation. It is regarded as relativistic
generalization of the nonrelativistic Schr\"odinger approach, or as
approximation to the homogeneous Bethe--Salpeter equation in its
instantaneous~limit. The nonlocality inherent to this kind of operators makes
hard to find analytical solutions. Nevertheless, rigorous analytical
statements can be proved by sophisticated methods \cite{TWR}:\begin{itemize}
\item Combining minimum--maximum principle and suitable operator inequalities
allows to derive ``semianalytical'' (or even analytical) upper bounds on all
energy levels~\cite{AUB}.\item Geometrical considerations summarized under
the term ``envelope technique'' yield ``semianalytical'' expressions for both
upper and lower bounds on eigenenergies \cite{ET}. For some particular
interactions these bounds can be represented in analytical form.\end{itemize}
Resulting eigenstates must be constructed numerically anyway, for instance,
by scanning the Hilbert space variationally using standard Rayleigh--Ritz
techniques or by integrating this equation of motion by conversion to some
equivalent matrix eigenvalue problem \cite{Num}. The achieved accuracy of
these approximate solutions can then be estimated by powerful criteria
derived from generalized --- in the present case relativistic --- virial
theorems \cite{RVT}.

\bibliographystyle{aipprocl}
\end{document}